 \documentclass[preprint2,longabstract]{aastex}

\slugcomment{Not to appear in Nonlearned J., 45.}

\shorttitle{Neutrinos and gamma-rays from nuclei in the jet}
\shortauthors{W. Bednarek}

\begin{document}


\title{High energy neutrinos produced in the accretion disks by neutrons from nuclei disintegrated in the AGN jets}


\author{W. Bednarek}
\affil{Department of Astrophysics, The University of Lodz, 90-236 Lodz, ul. Pomorska 149/153, Poland}

\email{bednar@uni.lodz.pl}



\begin{abstract}
We investigate the consequences of acceleration of nuclei in jets of active galaxies not far from the surface of an accretion disk. The nuclei can be accelerated in the re-connection regions in the jet and/or at the jet boundary, between the relativistic jet and its cocoon.
It is shown that the relativistic nuclei can efficiently fragment onto specific nucleons in collisions with the disk radiation. Neutrons, directed towards the accretion disk, take a significant part of energy from the relativistic nuclei. These neutrons develop a cascade in the dense accretion disk. We calculate the neutrino spectra produced in such hadronic cascade within the accretion disk. We propose that the neutrinos produced in such scenario from the whole population of super-massive black holes in active galaxies can explain the extragalactic neutrino background recently measured by the IceCube neutrino detector,
provided that $5\%$ fraction of galaxies is AGN and a few percent of neutrons reach the accretion disk. It is predicted that the neutrino signals in the present neutrino detectors, produced in terms of such a model, are not expected to be detectable even from the nearby radio galaxies similar to M87. 
\end{abstract}




%
%
\section{Introduction}

Hadrons are expected to be accelerated in the vicinity of super-massive black holes in active galactic nuclei (AGN) producing GeV-TeV $\gamma$-rays and also significantly contributing to the highest energy cosmic rays. The type of particles responsible for this $\gamma$-ray emission (leptonic or hadronic) can be only uniquely identified in the case of observations of neutrinos from these objects.

Recently a few tens of neutrino events have been detected by the IceCube
neutrino telescope in the TeV-PeV energy range (Aartsen et al. 2013, 2014). These events form a clear component above the atmospheric neutrino background. Their spectrum can be well described by a power law type extending up to a few PeV (Aartsen et al.~2015a). It has been proposed that these neutrinos are produced within active galaxies at cosmological distances (see for review e.g. Murase~2015 or Becker~2008). Different regions, such as AGN jets, AGN core regions or dense regions around super-massive black holes, have been suggested as possible emission sites.   
For example, in one of the first models of this type, Nellen et al.~(1993) proposed that protons accelerated in the jet interact with the accretion disk radiation. Neutrons, produced in such collisions, and also some of the protons, escape to the accretion disk producing neutrinos in hadronic collisions. This popular model for the neutrino production in blazars postulates interaction of the protons with the radiation field via pion production.
The radiation comes either from the jet, and/or from the accretion disk and/or is scattered around the jet in the broad line region (e.g. Mannheim~1995, Halzen \& Zas~1997, Bednarek \& Protheroe~1999, 
Atoyan \& Dermer~2001, M\"ucke \& Protheroe~2001, M\"ucke et al.~2003, Murase et al.~2014, Padovani et al.~2015). Most of these models predict neutrino spectra which usually flatten below $\sim$PeV energies due to the threshold of the pion production in 
photo-hadronic collisions. Therefore, such models have some difficulty in explaining the observed shape of the neutrino spectrum, which is close to the power law type between $\sim$10 TeV and a few PeV (as measured by the IceCube, see Fig.~3 in Murase~2015). The models which involve proton-proton interactions, as the dominant mechanism for neutrino production (e.g. Nellen et al.~1993, Beall \& Bednarek~1999, Schuster et al.~2002, Becker Tjus et al.~2014, Kimura et al.~2015, Hooper~2015), seems to be more plausible for modeling of the IceCube observations.  
For example, Becker Tjus et al. (2014) consider two scenarios for the neutrino production, i.e. acceleration of protons and their interaction with the matter of knots in the inner jets of FR-I galaxies and in the lobes of FR-II galaxies. The column density of matter of the order of $\sim$10$^{24\pm 1}$~cm$^{-2}$ is needed in the first scenario in order to explain the observed IceCube signal. The authors conclude that the second scenario requires a few orders of magnitude larger column density of the matter than expected in the radio lobes. Therefore, only first scenario can explain the IceCube results. 

In this paper we consider more complicated scenario in which produced neutrinos might significantly contribute to the flux observed by the IceCube. We note that a significant part of the matter which accretes onto the super-massive black hole (SMHB) in the AGN nucleus has to be composed from nuclei. This matter is likely to be more abundant in heavy nuclei, in respect to primordial matter containing $\sim$25$\%$ of helium, due to the efficient star formation occurring in the central parts of the parent galaxies. These nuclei are expected to be accelerated to relativistic energies in the re-connection regions in the jet and/or at the jet boundary layer. The
nuclei disintegrate in collisions with the accretion disk radiation producing relativistic neutrons
which, since neutral, can easily find the way to the dense accretion disk. Note that the neutrons from the disintegration of the nuclei have energies more than an order of magnitude lower than the neutrons 
produced in collisions of protons with photons due to the lower threshold on the first process. 
Therefore, neutrons from nuclei can easily produce multi-TeV neutrinos in consistency with the IceCube observations. We discuss the acceleration process of the nuclei and their interaction with the accretion disk in Sect.~2 and 3. The neutrino spectra produced in such a scenario are calculated in Sect.~4. The expected extragalactic neutrino background, calculated in terms of such a model, from the whole population of the SMBHs in the Universe, is compared with the recent observations by the IceCube 
telescope in Sect.~5. Finally, we discuss the observability of the neutrino signal from the example nearby active galaxy, M~87.

\section{Acceleration and dis\-in\-te\-gra\-tion of heavy nuclei in the jet}

We consider the processes in the inner part of the active galaxy in which an accretion disk forms around the SMBH. Jets in active galaxies are expected to be launched by a rotating black hole (e.g. Blandford \& Znajek~1977) or from the inner part of the accretion disk (e.g. Lovelace 1976, Blandford 1976). In fact, both processes mentioned above can play an important role. Therefore, the jet structure already at its base can be quite complicated. For example, the jet can be composed of a faster moving core region and a slower moving sheath.
The jet can be additionally surrounded by a cocoon. The cocoon can be formed by the matter surrounding the jet or by the outer layers of the jet expelled from the accretion disk, i.e. the accretion disk wind.
We consider two regions as responsible for the acceleration of particles in the jet/cocoon system, i.e. the re-connection regions in the jet and the transition region between the jet and the cocoon. We assume that an essential part of the jet power can be transferred to the relativistic nuclei in the inner part of the jet. Two mentioned above acceleration mechanisms of nuclei are discussed in a more detail below. As a result of the disintegration of nuclei in collisions with the disk radiation, relativistic neutrons are extracted. 
In fact, neutrons can take a significant part of the energy of the accelerated nuclei, between $\sim$$8^{-1}$ (in the case of the presence of only primordial Helium, assuming the primordial He abundance of $\sim$0.25, e.g. Peimbert et al.~2007) and up to $\sim$$2^{-1}$ (if accelerated hadrons are dominated by heavy nuclei which might appear due to the nuclear burning within the stars which form the central stellar cluster around SMBH). These neutrons irradiate the accretion disk interacting with a large column density of matter. The schematic representation of the processes discussed in this paper are shown in Fig.~1.

\begin{figure}
\vskip 6.truecm
\includegraphics{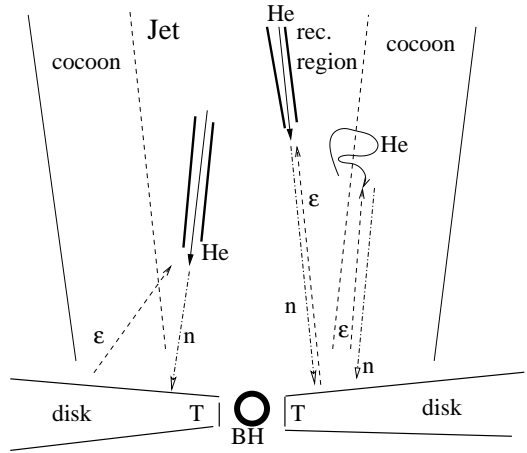}
\caption{Schematic presentation of the inner part of the jet surrounded by a slowly moving cocoon. Nuclei (e.g. He) are expected to be accelerated either in the re-connection regions within the relativistic jet or at the boundary between the fast jet and its slow cocoon. We assume that nuclei, accelerated in the re-connection regions, are directed towards the accretion disk. Nuclei 
can be also accelerated at the boundary layer between the jet and the cocoon. In such a case, they are expected to spend half of their residence time in the cocoon. In the cocoon frame, nuclei have close to isotropic distribution, being directed towards the accretion disk for a significant amount of time. Relativistic nuclei disintegrate in collisions with the accretion disk radiation onto specific nucleons. Neutrons (with large Lorentz factors) reach dense regions of the accretion disk before decaying into protons. They produce high energy $\gamma$-rays and neutrinos in collisions with the matter of the accretion disk.}
\label{fig1}
\end{figure}
\subsection{Re-connection regions}

We assume that nuclei can be accelerated in re-connection regions, which are oriented in the general direction towards the accretion disk in the SMBH rest frame, to energies allowing them efficient disintegration into individual nucleons. 
Such situation is expected when the re-connection process is driven by the perpendicular component of the magnetic field in the jet.
Acceleration of particles in the re-connection regions has been considered as an important process for energization of particles in different astrophysical scenarios (e.g. see recent reviews by Kagan et al.~2015 and Uzdensky ~2016). It is expected to be responsible for the acceleration of particles in relativistic jets of blazars (see e.g. Romanova \& Lovelace~1992, Lesch \& Birk~1998, Larrabee et al.~2003, Lyutikov~2003, Jaroschek et al.~2004, Giannios et al.~2009, Sironi \& Spitkovsky~2014).
For example, following Giannios (2010), we assume that nuclei within the jet can be accelerated in the re-connection regions. 
We apply a simple model for the acceleration process in the re-connection region.
Its length, $L_{\rm rec}$, scales with the distance from the base of the jet (e.g. see Bednarek et al.~1995). The distance along the jet, $R$, is given in units of the inner radius of the jet, $R_{\rm in}$, according to $r = R/R_{\rm in}$. Then, $L_{\rm rec}$ is linked to the SMBH mass,
\begin{eqnarray}
L_{\rm rec} = \xi R_{\rm in}r\approx 10^{14}\xi_{-1}M_9r~~~{\rm cm},
\label{eq1}
\end{eqnarray}
\noindent
where $R_{\rm in} = 3R_{\rm s}\approx 9\times 10^{14}M_9$ cm is the inner radius of the jet, 
$R_{\rm S}$ is the Schwarzschild radius of the black hole,  $M_{\rm BH} = 10^9M_9$ M$_\odot$ is the mass of the black hole in the Solar masses, $\xi = 0.1\xi_{-1}$ is the scaling factor of the re-connection region assumed to be comparable to the perpendicular extend of the jet which is $\sim$0.1r for the opening angle of the order of $\sim$0.1 rad.

We estimate the mean free path for the disintegration of the Helium nuclei (expected to be the most abundant between the heavy nuclei) in the radiation field of the accretion disk,
\begin{eqnarray}
\lambda_{\rm des}^{\rm He} = (n_{\rm ph}\sigma_{\rm des}^{\rm He})^{-1}\approx 9.4\times 10^{11}r^2/T_{4.5}^3~~{\rm cm},
\label{eq2}
\end{eqnarray}
\noindent
where $n_{\rm ph}\approx 5.3\times 10^{14}T_{4.5}^3/r^2$~ph.~cm$^{-3}$ is the density of the diluted, by the factor $r$, black body radiation with the temperature at the inner disk radius $T_{\rm D} = 3\times 10^4T_{4.5}$ K, $\sigma_{\rm des}^{\rm He}\approx 2\times 10^{-27}$ cm$^2$ is the cross section for the disintegration process of He nuclei at its maximum (see Appendix A in Cyburt et al. 2003).
As estimated in Section~4, the surface temperature in the inner part of an accretion disk around a massive black hole is expected to be of the order of a few $10^4$ K. This is consistent with the observations of the ultraviolet excess in the spectra of some quasars (e.g. $(2-3)\times 10^4$ K in the case of 3C 273, Malkan \& Sargent~1982).

In order to provide an efficient disintegration of nuclei, $\lambda_{\rm des}^{\rm He}$ should be shorter than the size of the re-connection region. This condition is fulfilled for a distance from the base of the jet (obtained from the comparison of Eq.~1 and 2),
\begin{eqnarray}
r < 100\xi_{-1}M_9T_{4.5}^3.
\label{eq3}
\end{eqnarray}
Therefore, it is expected that the nuclei can disintegrate inside the inner jet,
i.e. within $\sim$pc distance scale. 

In contrary, helium nuclei have to be accelerated to large enough energies in order to suffer efficient disintegration. The electric field strength within the re-connection region is parametrised by,
\begin{eqnarray}
V_{\rm rec} = \eta cB\approx 3\times 10^3\eta_{-1}B_2 r^{-\beta}~~{\rm V~cm^{-1}},
\label{eq4}
\end{eqnarray}
\noindent
where the magnetic field strength along the jet is described by $B = B_{\rm in}r^{-\beta}$ (with $\beta = 1$ for the toroidal component and $\beta = 2$ for the longitudinal component), the magnetic field at the base of the jet is $B_{\rm in} = 100B_2$ G, $\eta = 0.1\eta_{-1}$ is the efficiency of the re-connection process, and $c$ is the velocity of light. It has been recently shown that if the re-connection process occurs in the electron-proton plasma, the acceleration efficiency can reach values as large as $\sim$0.2 (see Sironi, Petropoulou \& Giannios~2015). 
The maximum Lorentz factors of the nuclei can be estimated from, 
\begin{eqnarray}
E_{\rm He}^{\rm max} & = & V_{\rm rec}L_{\rm rec}Z_{\rm He} \cr
& \approx & 3\times 10^{17} \eta_{-1}\xi_{-1}M_9B_2 Z_{\rm He}r^{1 -\beta}~~{\rm eV}.
\label{eq5}
\end{eqnarray}
\noindent
where $Z_{\rm He} = 2e$ is the charge of He nuclei and $e$ is the elementary charge.
Note however, that the inner jet might be already mildly relativistic with the bulk Lorentz factor, $\Gamma_{\rm j}$,  of the order of a few
(see e.g. Vlahakis \& Koenigl~2004). Then, the maximum energies of the nuclei in the disk reference frame are reduced by the value of this Lorentz factor. On the other hand, energies of nuclei can be also larger than estimated 
in Eq. (5) provided that the BH rotates fast since in such a case the accretion disk should extend closer to the BH and the magnetic field in the inner jet should be stronger than in the case of the Schwartzschild BH. Therefore, the above formula give only an order of magnitude estimate of energies to which the nuclei can be accelerated.

The He nuclei are efficiently disintegrated when the soft radiation from the accretion disk reaches energies above $E^{\rm He}_{\rm min}\sim$20 MeV in their reference frame (e.g. see Appendix A in Cyburt et al.~2003). This condition is met for the energies of the nuclei of the order of, 
\begin{eqnarray}
E_{\rm He}^{\rm min} = {{m_{\rm p}c^2A_{\rm He}E^{\rm He}_{\rm min}}\over{3k_{\rm B}T_{\rm D}}}
\approx {{2.4\times 10^{15}A_{\rm He}}\over{T_{4.5}}}~~{\rm eV},
\label{eq6}
\end{eqnarray}
\noindent
where $A_{\rm He} = 4$ is the mass number of He nuclei, $m_{\rm p}c^2$ is the proton rest energy, $c$ is the velocity of light, and $k_{\rm B}$ is the Boltzmann constant.
Note that neutrons from the disintegration of the nuclei are expected to have Lorentz factors similar to those of the  parent nuclei. The obvious condition, i.e. $E_{\rm He}^{\rm max} > E_{\rm He}^{\rm min}$, is met in the case of $\beta = 2$ in the inner region of the jet, i.e within the distance from the jet base,
\begin{eqnarray}
r < 1.2\times 10^2 \eta_{-1}\xi_{-1}M_9B_2T_{4.5}Z_{\rm He}/A_{\rm He}.
\label{eq7}
\end{eqnarray}
\noindent
The conditions given by Eq.~(3) and Eq.~(7) postulate that the reconnection process has to occur already in the
inner part of the jet. In fact, the theory of this process in the AGN jets predicts that the dissipation region 
is located at the distance $R_{\rm diss}\approx \Gamma_{\rm j}^2R_{\rm S}/\varepsilon\approx 40\Gamma_2^2R_{\rm S}/\varepsilon_{-1}$, where $\Gamma_{\rm j} = 2\Gamma_2$ is the jet Lorentz factor and $\varepsilon = 0.1\varepsilon_{-1}$ is the reconnection speed (see Eq.~1 in Giannios~2013). Therefore, for mildly relativistic jets the acceleration of nuclei in the reconnection process seems to be present already in the inner jet.

If the structure of the magnetic field in the jet is better described by $\beta = 1$, then the above condition is met everywhere in the jet provided that the magnetic field strength at the base of the jet is,
\begin{eqnarray}
B_{\rm in} > 0.8 A_{\rm He}/(\eta_{-1}\xi_{-1}M_9T_{4.5}Z_{\rm He})~~~{\rm G}.
\label{eq8}
\end{eqnarray}
\noindent
Note that the conditions for the disintegration of nuclei in jets of active galaxies can be 'easier' fulfilled (i.e. for weaker magnetic fields) for the SMBHs with larger masses.

We assume that a part of neu\-trons lib\-er\-ated from nuc\-lei im\-pinge onto the ac\-cre\-tion disk. 
Neutrons with the Lorentz factors above $\gamma_{\rm n} > E_{\rm He}^{\rm min}/(A_{\rm He}m_{\rm p})$, can travel characteristic distances of the order of $X_{\rm n} > c\tau_{\rm n}\gamma_{\rm n} > 8\times 10^{18}$ cm. This distance scale is clearly larger than the distance scale for the efficient extraction of neutrons from the nuclei (given by Eq.~7) for the black holes with masses $\le 10^{10}$ M$_\odot$. Therefore, we conclude that in most cases the neutrons extracted from the nuclei, moving towards the accretion disk, can reach the disk surface before decaying.

We note that extraction of neutrons from He nuclei is more important process than the energy losses of the nuclei on the Bethe-Heitler $e^\pm$ pair production in collisions with the disk radiation. 
Although the cross section for the disintegration of He nuclei is similar to 
the cross section for the Bethe-Heitler e$^\pm$ pair production, the in-elasticity coefficient for energy extraction by dissolved nucleons from the He nuclei ($\sim$m$_{\rm n}$/m$_{\rm He}$) is about three orders of magnitude larger 
than for the $e^\pm$ pair production ($\sim$2m$_{\rm e}$/m$_{\rm He}$, e.g. Chodorowski, Zdziarski \& Sikora 1992), where m$_{\rm e}$, m$_{\rm n}$, and m$_{\rm He}$ are the rest masses of the lepton, neutron and He nuclei, respectively.  Therefore the process of Bethe-Heitler $e^\pm$ pair production can be neglected as the energy loss process in respect to the energy loss process of the disintegrated He nuclei.

Similar constraints can be also obtained in the case of production of neutrons in collisions of protons with the disk radiation.
However since the cross section for the photo-pion production process, $p-\gamma\rightarrow n + \pi$, is about an order of magnitude lower and the energy threshold for the pion production is about of an order of magnitude larger than for efficient disintegration of a He nuclei, the constraints on neutron production by protons are at least an order of magnitude more restrictive. Therefore, the neutron production in p-$\gamma$ collisions is not expected to be effective process of neutrino production. Moreover, neutrons produced in this last process are expected to have clearly larger energies than required for efficient production of neutrinos with the simple power law spectra extending between $\sim$10 TeV and a few PeV.

\subsection{Jet boundaries}

Essential role in the collimation of a jet plays the surrounding matter and/or the wind from the accretion disk. Therefore, jets are expected to have regions (boundaries) filled with
plasma moving with various velocities.
In fact, recent observations of the inner jet structure in the nearby radio galaxy Cygnus A show that the transverse width of the jet, already very close to the jet base, is significantly larger than the radius of the innermost stable orbit of the super-massive black hole (Boccardi et al.~2016). This suggests that the accretion disk is contributing to the jet launching. The jet likely have a faster inner section, powered by the black hole or the inner disk, and a slower outer section anchored in the more distant parts of the disk. The border between these two parts of the jet has been proposed to provide good conditions for the acceleration of particles.
In fact, the acceleration of particles in the plasma at the shear flows has been studied since the works by Berezhko \& Krymskii~(1981 and Berezhko~(1981). The case of the boundary between relativistic jet and its cocoon was discussed by Ostrowski (1990, 1998, 2000) and also Bisnovatyi-Kogan \& Lovelace~(1995). These mechanisms of particle acceleration at the jet boundary are expected to produce relativistic particles with a flat power law spectra (e.g. Ostrowski~1990,1998,2000, Rieger \& Duffy~2006).  
Nuclei accelerated in such mechanism should spend significant time within the cocoon of the jet in which their distribution is close to isotropic. Therefore, they can preferably interact with the nearby accretion disk radiation. As a result, the nuclei can lose nucleons in the photo-disintegration process. We expect that a part of these neutrons released in the disintegration process of the nuclei (as an example estimated in Sect.~5) propagate towards the accretion disk and interact with the disk matter. These neutrons will initiate the cascade in the optically thick accretion disk. Neutrinos, produced in hadronic interactions, escape from the disk without absorption. On the other hand, $\gamma$-rays are expected to be absorbed in the disk due to the interactions with the disk matter and radiation field. 

The simulation of the particle acceleration process in such conditions by Ostrowski~(1990) show that in favorable conditions considered acceleration process can be very rapid. The acceleration length in the observer's frame, due to the particle advection along the jet flow, can be
$L_{\rm acc}\sim \alpha_{\rm o} r_{\rm L}$, where $r_{\rm L}\approx 6\times 10^6\gamma_{\rm He}B_{\rm G}^{-1}$ cm is the Larmor radius of the nuclei and the parameter $\alpha_{\rm o}$ can be as small as $\sim$10. We are interested in situations in which the accelerated nuclei are efficiently disintegrated. Then, the acceleration length of nuclei, 
\begin{eqnarray}
L_{\rm acc} & \sim & 6\times 10^6\alpha_{\rm o} \gamma_{\rm He}B_{\rm G}^{-1} \cr
& \sim & 1.8\times 10^{10}\alpha_{\rm o} T_{4.5}r^\beta B_2^{-1}~~~{\rm cm}, 
\label{eq9}
\end{eqnarray}
should be at least equal (or shorter) than the mean free path for their disintegration
(see Eq.~2), provided that on this distance scale the nuclei reach energy above the threshold for disintegration (given by Eq.~6). The condition, $L_{\rm acc} \le \lambda_{\rm des}^{\rm He}$, is fulfilled for the distance from the base of the jet,
\begin{eqnarray}
r\ge 0.02\alpha_{\rm o} T_{4.5}^4B_2^{-1},
\label{eq10}
\end{eqnarray}
\noindent
for the case of the dominant toroidal structure of the magnetic field in the jet, i.e. $\beta = 1$ and
$\alpha_{\rm o}\sim 10$. 
If the longitudinal magnetic field component dominates within the jet, i.e. $\beta = 2$, then the condition, $L_{\rm acc} \le \lambda_{\rm des}^{\rm He}$, implies the magnetic field at the base of the jet,
\begin{eqnarray}
B_{\rm in}\ge 2\alpha_{\rm o} T_{4.5}^4~~~{\rm G}.
\label{eq11}
\end{eqnarray}
\noindent
The conditions, given by Eq.~(10) and (11), are consistent with the maximum distance from the base of the jet for which the disintegration process can become efficient (given by Eq.~3). 

We estimate the maximum Lorentz factors of nuclei, accelerated at a specific distance, L, from the base of the jet,  in the jet boundary layer by comparing their characteristic acceleration length, $L_{\rm acc}$,  with the distance scale along the jet, $L\approx 10^{15}M_9r$ cm, 
\begin{eqnarray}
\gamma_{\rm He}^{\rm max}\approx 1.5\times 10^{10}M_9B_2r^{1-\beta}/\alpha_{\rm o}.
\label{eq12}
\end{eqnarray}
\noindent
Assuming $\alpha_{\rm o}\sim 10$, this maximum Lorentz factor is $\gamma_{\rm He}^{\rm max}\approx 1.5\times 10^9M_9B_2$
for $\beta = 1$ and $\gamma_{\rm He}^{\rm max}\approx 1.5\times 10^9M_9B_2/r$ for $\beta = 2$. These maximum energies of nuclei are clearly above the minimum energies required for their efficient disintegration in the disk radiation field (see Fig.~6).

The mechanism of particle acceleration at the jet boundary described above is expected to produce relativistic particles with a flat power law spectrum (e.g. Ostrowski~1990,1998,2000, Rieger \& Duffy~2006).  We conclude that the nuclei, accelerated in the jet not far from the accretion disk, can suffer efficient disintegration process. As a result, relativistic neutrons are injected towards the dense accretion disk. The neutrons reach the disk producing neutrinos as a decay products of charged pions. Our aim is to estimate the contribution of these neutrinos to the high energy extragalactic neutrino background in the Universe.

\section{Interaction of neutrons with an accretion disk}

We assume that a part of neutrons, extracted from the relativistic nuclei in the jet region, are directed towards the accretion disk. These neutrons have large enough Lorentz factors to reach the accretion disk before decaying.
Neutrons have the power law spectrum consistent with the spectrum of the accelerated nuclei. 
Their Lorentz factors are equal to Lorentz factors of their parent nuclei. 

The accretion disks around SMBHs in radio loud active galaxies are well described by the Shakura-Sunyaev (1973) disk model, i.e. they are optically thick and geometrically thin. Such type of disks form provided that the accretion rate is not very far from the critical Eddington accretion rate. The surface mass density in the inner part of such a geometrically thin Shakura-Sunyaev (1973) type disk is, 
\begin{eqnarray}
\Sigma (r) = 4.6\alpha^{-1}\dot{m}^{-1} r^{3/2} (1 - r^{-1/2})^{-1}~~{\rm g~cm^{-2}},
\label{eq13}
\end{eqnarray}
\noindent
(see Eq. 2.9 in Shakura \& Sunyaev~1973), where $r$ is the distance from the black hole expressed in units of the inner radius of the accretion disk, $r_{\rm in} = 3r_{\rm S} = 6GM_{\rm BH}/c^2$,
$\dot{m}$ is the accretion rate in units of the Eddington accretion rate, $\alpha$ is the viscosity coefficient, $G$ is the gravitational constant. The above formula is valid for distances, $r < 150 (\alpha m)^{2/21}\dot{m}^{16/21}$, where $m = M_{\rm BH}/M_\odot$. For reasonable values of the viscosity parameter, $\alpha = 0.1$, the accretion rate $\dot{m} = 0.1$,
and the mass of the black hole $m = 10^9$, the formula is valid for $r < 150$.
Density of the matter in the inner disk with such parameters lays between $4.4\times 10^3$ g cm$^{-2}$ for r = 2 and $8.5\times 10^5$ g cm$^{-2}$ for $r = 150$.  On the other hand, the density of matter in the disk is low since the thickness, $z$, of the disk around the super-massive black hole is quite large (see Eq.~2.8 in Shakura \& Sunyaev~1973), 
\begin{eqnarray}
z(r)\approx 3.2\times 10^6\dot{m} m (1 - r^{-1/2})~~{\rm cm}. 
\label{eq14}
\end{eqnarray}
\noindent
For the parameters, $\dot{m} = 0.1$ and $m = 10^9$, the half thickness of the disk is $z\approx 9.4\times 10^{13}$ cm. 
The collision length of neutrons on the hadronic interactions in the inner disk can be estimated from 
$\lambda_{\rm np} = (n(r)\sigma_{\rm np})^{-1}\approx 1.5\times 10^{12}$~cm, where 
$n(r) = \Sigma(r) N_{\rm A}/\mu_{\rm A} z(r)\approx 2.3\times 10^{13}$ cm$^{-3}$, $N_{\rm A} = 6\times 10^{23}$~mol$^{-1}$
is the Avogadro number, $\mu_{\rm A} = 16/13$~g~mol$^{-1}$ is the average molar mass of the matter composed of $75\%$ of hydrogen and $25\%$ of helium, and $\sigma_{\rm np} = 3\times 10^{-26}$ cm$^{-2}$ is the cross section for hadronic collisions of neutrons with the matter. This interaction length is much shorter than the thickness of the accretion disk.

Density of the matter in the disk is a few orders of magnitudes lower than the density of the Earth's atmosphere. Therefore, pions and muons with considered energies, produced in the interactions of neutrons with the matter, decay onto neutrinos before interacting with the matter. On the other hand, $\gamma$-rays are effectively converted into 
$e^\pm$ pairs in the interactions with the Coulomb field of the nuclei and with the soft 
radiation within the accretion disk. The optical depths for these processes are 
much larger than unity for the characteristic disk temperatures of the order of
$3\times 10^4$ K and the column density of the matter and the thickness of the accretion disk estimated above.  
Note that, some of the neutrons during such hadronic interactions convert to protons. These protons are efficiently captured and isotropised by the magnetic field in the disk. 
Their Larmor radii, $R_{\rm L}\approx 3\times 10^{12} E_{\rm PeV}/B_{\rm G}$ cm (where proton energy $E = 1E_{\rm PeV}$ PeV), are expected to be  smaller than the depth of the first interaction of neutron in the inner part of the disk, $d\approx 0.01 z\approx 10^{12}$ cm (see above). This condition is fulfilled for protons with energies below, 
$E\approx 30B_{2}$ PeV, where $B_{\rm G}$ is the magnetic field strength in the disk. 
We estimate the magnetic field in the disk by assuming that
its energy density is comparable to the energy density of disk radiation field, i.e. $B\approx 370T_{4.5}$~G. Then, protons with energies below $\sim$100 PeV are isotropised. Due to the presence of the magnetic field in the disk, neutrinos, produced during the cooling process of the primary neutrons, are expected to be emitted quasi-isotropically from the inner disk.

\section{Spectra of neutrinos}

\begin{figure*}
\vskip 5.2truecm
\includegraphics{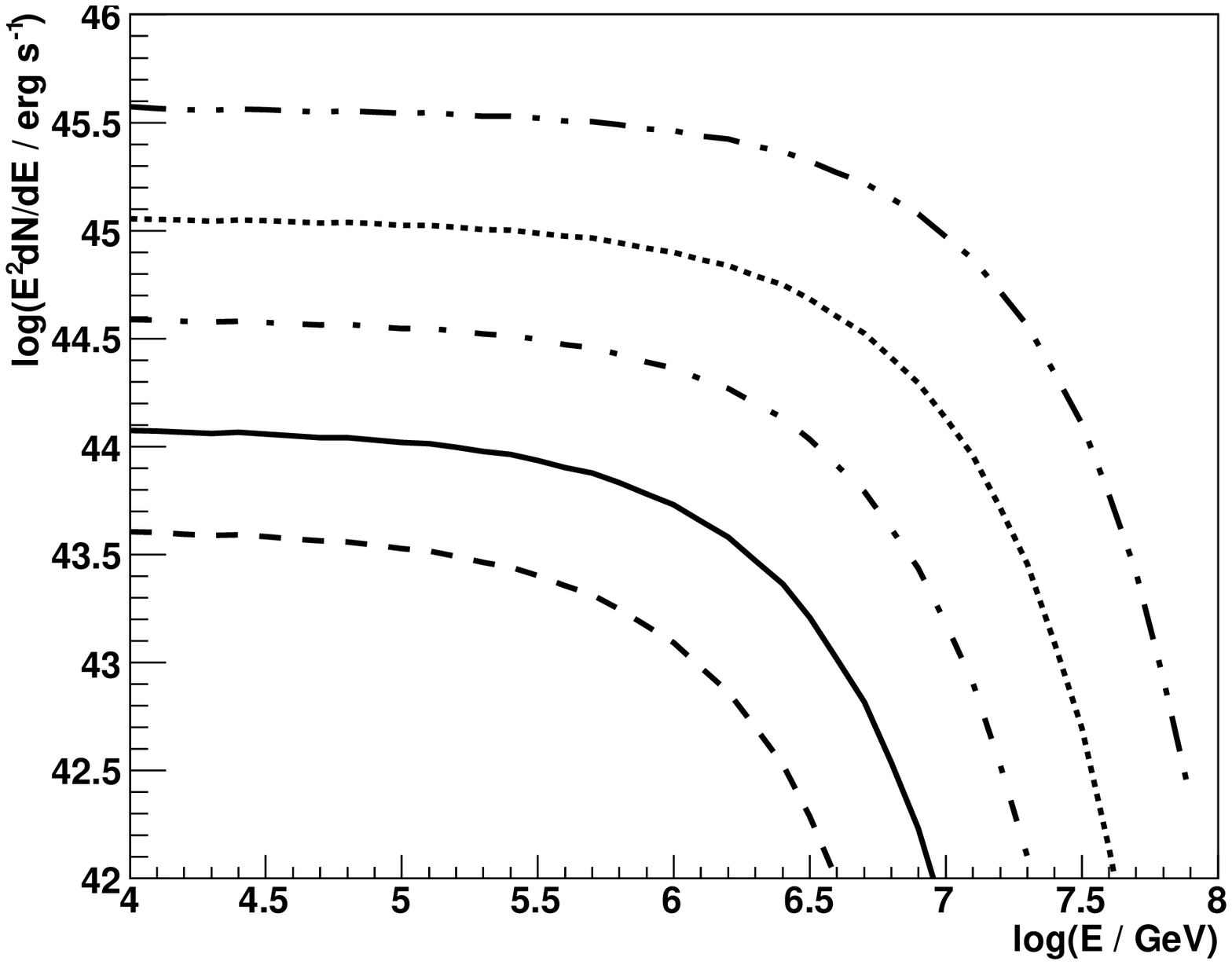}
\includegraphics{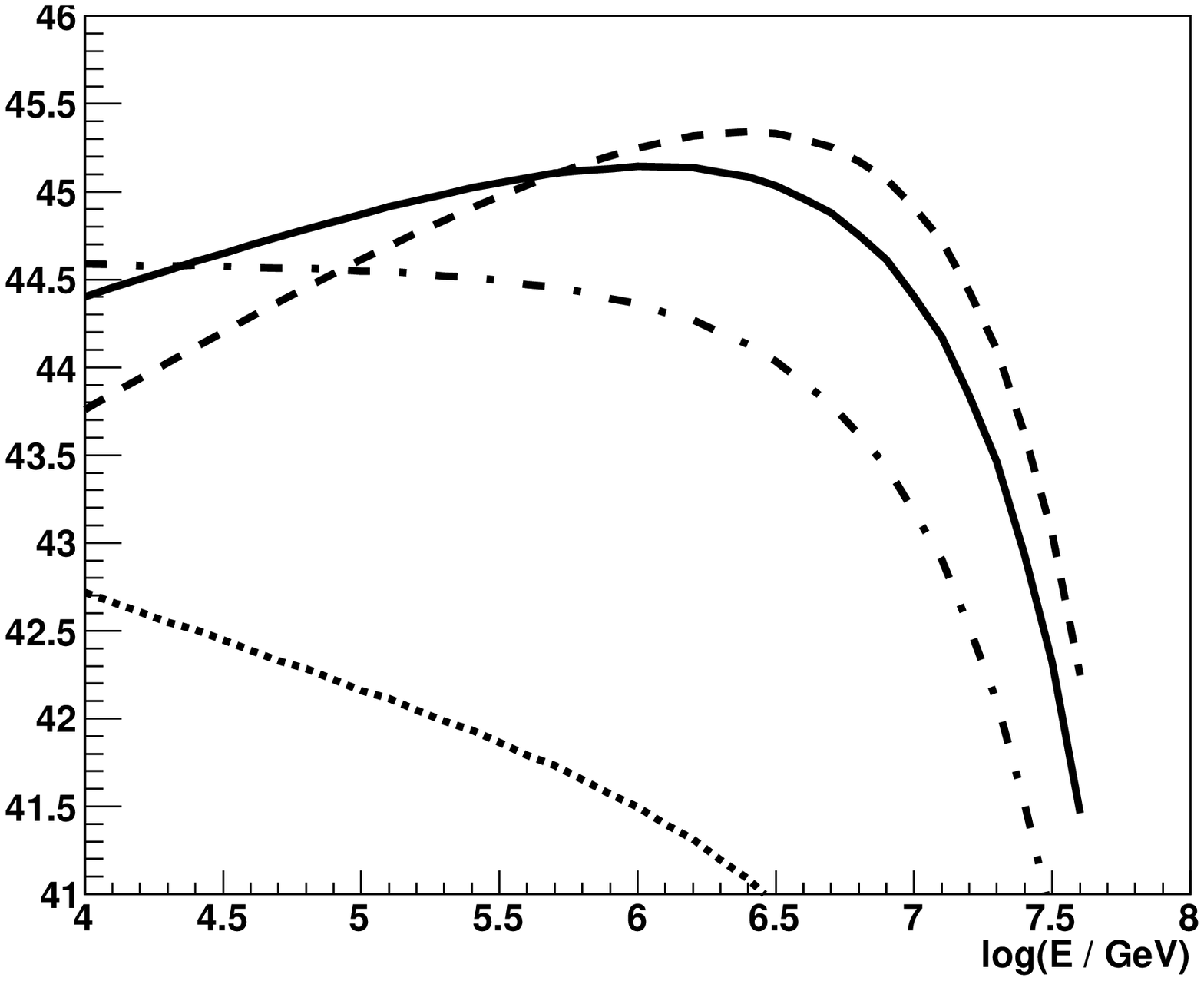}
\includegraphics{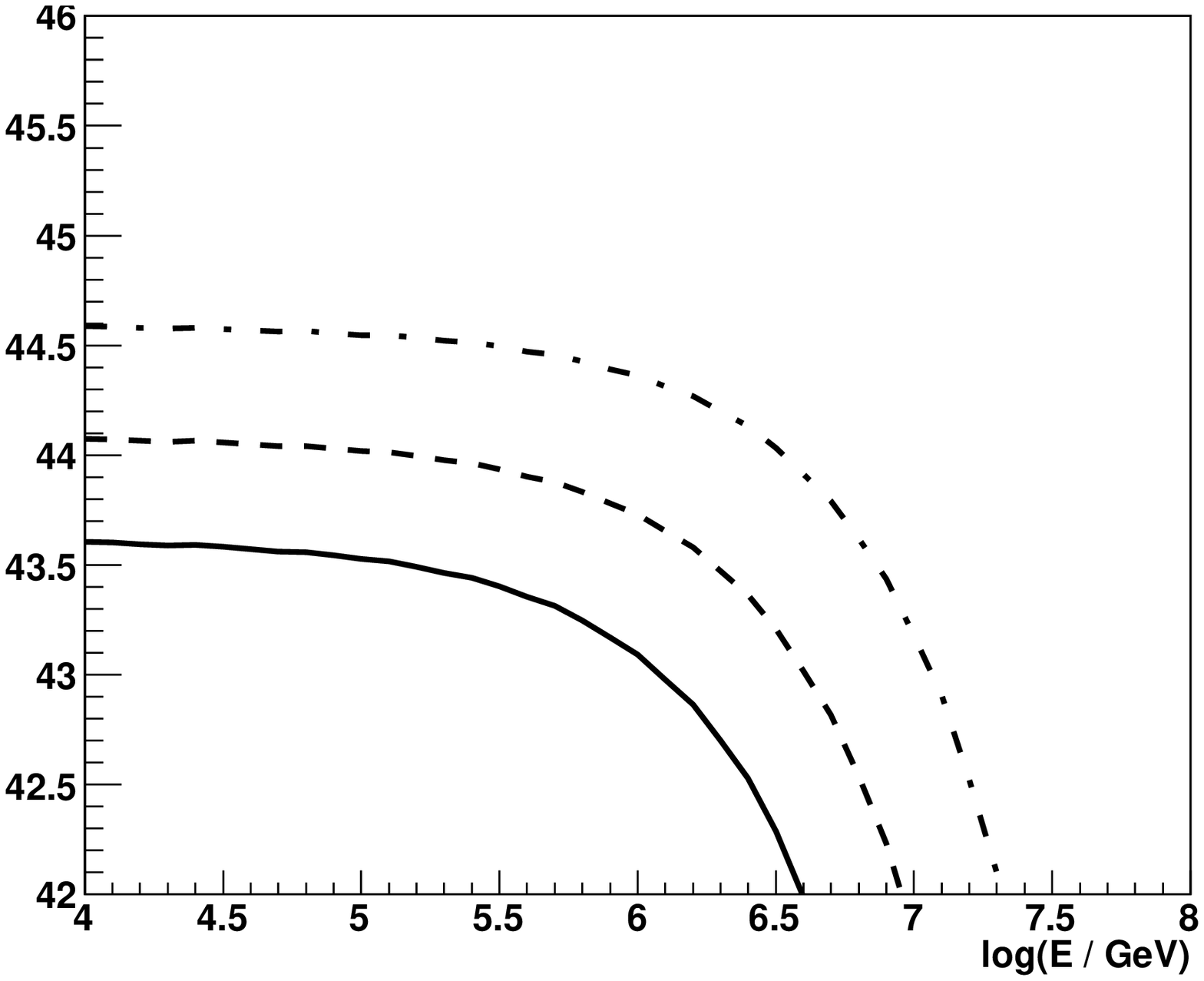}
\caption{Neutrino (all flavor) spectra produced by neutrons in collisions with the matter of the optically thick (and geometrically thin) Shakura-Sunyaev type accretion disk for different masses of the supermassive black hole $M_{\rm BH} = 10^{10}$ M$_\odot$ (dot-dot-dashed), $3\times 10^9$ M$_\odot$ (dotted), $10^9$ M$_\odot$ (dot-dashed), $3\times 10^8$ M$_\odot$ (solid) and $10^8$ M$_\odot$ (dashed) (on the left). The neutrons have a power law spectrum with spectral index equal to $\delta = 2$ and an exponential cut-off (see Eq.~15) at the energy given by~Eq.~20. 
The neutron spectrum is calculated assuming that the jet takes half of the gravitational energy of the accreting matter. The black hole accretes the matter with the efficiency equal to $\chi = 0.1$ of the Eddington accretion rate. $10\%$ of the jet power is transferred to relativistic He nuclei. The neutrino spectra for different spectral indexes of neutrons, equal to $\delta = 1$ (dashed curve), 1.5 (solid), 2.0 (dot-dashed), and 2.5 (dotted), are shown in the middle figure. The SMBH mass is fixed on $10^9$ M$_\odot$ and the other parameters are as defined above. The spectra for different efficiency of the accretion process, $\chi = 0.1$ (dot-dashed curve), 0.03 (dashed), and 0.01 (solid), are shown on the right figure.}
\label{fig2}
\end{figure*}

We calculate the all flavor neutrino spectra produced in the above discussed scenario for the SMBHs with different masses. As an example, we assume that nuclei are accelerated with the power law spectrum and an exponential cut-off. Neutrons are extracted from these nuclei with a similar spectrum. It is of the form,
\begin{eqnarray}
dN_{\rm n}/dE_{\rm n}\propto E_{\rm n}^{-\delta}\exp(-E_{\rm n}/E_{\rm n}^{\rm max}).
\label {eq15}
\end{eqnarray}
\noindent
The power taken by neutrons (i.e. the normalization factor in this spectrum), can be obtained by relating it to the accretion luminosity of the SMBH. This accretion luminosity can be again related to the Eddington luminosity of the SMBH,
assuming the cosmological redshift averaged accretion efficiency, 
\begin{eqnarray}
L_{\rm acc} = \chi L_{\rm Edd} = 1.3\times 10^{46}\chi_{-1}M_9~~{\rm erg~s^{-1}}, 
\label{eq16}
\end{eqnarray}
where $L_{\rm Edd}$ is the Eddington luminosity, and the efficiency of the accretion process is $\chi = 0.1\chi_{-1}$. In the case of the radio-loud AGN this value has been estimated to be in the range $\sim$(0.01-0.1) (e.g. Wu \& Liu~2004).
It is assumed that approximately a half of the energy released by accretion is irradiated from the accretion disk surface, $L_{\rm D} = 0.5L_{\rm acc}$, and a half of the accretion energy is transferred to the jet, 
$L_{\rm j} = 0.5L_{\rm acc}$. Assuming the Shakura-Sunyaev~(1973) disk model, we estimate the characteristic temperature of the radiation emitted from the inner disk on
\begin{eqnarray}
T_{\rm D} = \left({{L_{\rm D}}\over{4\pi r_{\rm in}^2\sigma_{\rm SB}}}\right)^{1/4}\approx 
5.5\times 10^4{{\chi_{-1}^{1/4}}\over{M_9^{1/2}}}~~~{\rm K}.
\label{eq17}
\end{eqnarray}
The inner disk temperature determines the minimum energy of He nuclei for which their disintegration in the disk radiation becomes efficient (Eq.~6). In order to estimate the characteristic maximum energies of neutrons, $E_{\rm n}^{\rm max}$, as a function of the SMBH mass, we set the upper limit on the magnetic field strength at the base of the jet by
assuming that the jet power is mainly curried out in the form of the Poynting flux,
\begin{eqnarray}
L_{\rm j}\approx \pi r_{\rm in}^2 c (B_{\rm in}^2/8\pi)^2.
\label{eq18}
\end{eqnarray}
This relation allows to estimate the magnetic field on,
\begin{eqnarray}
B_{\rm in}\approx 1.3\times 10^3(\chi_{-1}/M_9)^{1/2}~~~{\rm G.}
\label{eq19}
\end{eqnarray}
Introducing this value of the mag\-ne\-tic field strength to the formula for the maximum energies to which nuclei can be accelerated in the jet (see Eqs.~5 and ~12) and applying the assumed scaling values for other parameters, we estimate the characteristic maximum energies of neutrons on,
\begin{eqnarray}
E_{\rm n}^{\rm max} = 2\times 10^9(\chi_{-1}M_9)^{1/2}~~~{\rm GeV.}
\label{eq20}
\end{eqnarray}
For simplicity, in Eq.~5 and Eq.~12, we fixed the parameter describing the distribution of the magnetic field in the jet on $\beta = 1$. Then, the maximum energies of the nuclei are independent on the distance from the SMBH. On the other hand, the minimum energies of neutrons are obtained from the combination of Eq.~6 and Eq.~17, 
\begin{eqnarray}
E_{\rm n}^{\rm min} = 1.4\times 10^6M_9^{1/2}/\chi_{-1}^{1/4}~~~{\rm GeV.}
\label{eq21}
\end{eqnarray}
They depend on the mass of the SMBH and on the accretion efficiency of the SMBHs through the dependence on the temperature in the inner part of the accretion disk.

A part of the neutrons, extracted from the nuclei, move towards the accretion disk and interact with the matter producing neutrinos in hadronic collisions.  As an example, we show the dependence of the neutrino spectra on the SMBH mass (see Fig.~2 on the left), on the spectral index of the spectrum of nuclei (see Fig.~2 in the middle), and different efficiency of the accretion process (see Fig.~2 on the right). Note that the spectra of particles with indexes as low as $\delta = 1$ are expected in the case of the re-connection process (e.g. Larrabee et al. ~2003).
The spectra shown in the first two figures on Fig.~2, are calculated for the SMBH accreting the matter at the rate of $\chi = 10\%$ of the Eddington accretion rate. It is assumed that half of this accretion energy is taken by the jet. The nuclei take $10\%$ of the jet power.

The power in the neutrino spectrum, and its high energy cut-off, depends in a simple way on the mass of SMBH and the acceleration efficiency.  In the considered range of SMBH masses, the neutrino spectra clearly extend up to the PeV energies for the efficiency of acceleration close to $\chi = 0.1$. Such SMBHs might be responsible for the PeV neutrino events recently detected by the IceCube telescope (see Aartsen et al.~2013). However, the spectral index of accelerated nuclei has to be close to 2 in order to transfer significant energy into the PeV neutrinos. If the accretion efficiency is closer to $\sim$0.01, as obtained by Wu \& Liu~(2004) for closer radio loud AGNs, the spectral indexes of accelerated nuclei should be flatter than 2 in order to efficiently produce $\sim$PeV neutrinos.

In the next section, we estimate the extragalactic neutrino background (ENB) produced by the population of accreting SMBHs in the Universe. The neutrino spectra expected in our model are compared with the recent observations of the neutrino events by the IceCube telescope.

\section{Extragalactic neutrino background}

We wonder whether neutrinos produced in the scenario described above can explain the observations of the very high energy Extragalactic Neutrino Background (ENB) reported recently by the IceCube Collaboration. In order to determine the contribution of the neutrinos, produced in the accretion disks around SMBHs, to the ENB, we
integrate the neutrino spectra over the luminosity function of the spheroids around SMBHs in active galaxies. This luminosity function is related to the masses of SMBHs. We integrate this formula over the part of the Universe up to the redshift $z = 2$. 

The diffuse neutrino flux is then given by,
\begin{eqnarray}
\Phi_\nu & = & {{c}\over{4\pi H_{\rm o}}}  \int^{\rm z_{\rm max}}_{0}{{dz}\over{[(1 + z)^3\Omega_{\rm m} + \Omega_\Lambda]^{1/2}}} \cr
 &  & \int dL {{dN(L,z)}\over{dLdV}} {{dN_\nu(E_\nu(1 + z))}\over{dE_\nu dt}},
\label{eq22}
\end{eqnarray}
\noindent
where $dN(L,z)/dLdz = (dN(L,z)/dLdV)(dV/dz)$ is the spheroid luminosity, $L$, function, $z$ is the redshift, and $V$ is the volume. The spheroid luminosity is related to the SMBH mass through the formula, 
\begin{eqnarray}
\log{{M_{\rm BH}}\over{M_\odot}} & =  & 1.13\log{{L}\over{L_\odot}} - 4.11 \cr
&  & - 0.316z + 1.4\log(1 + z),
\label{eq23}
\end{eqnarray}
taken from Li et al.~(2011). 
The luminosity function is expressed as a Schechter (1976) function,
\begin{eqnarray}
{{dN(L,z)}\over{dLdV}} = {{\Phi_0(z)}\over{L_\star}} \left({{L}\over{L_\star}}\right)^{-1.07}
\exp(-{{L}\over{L_\star}}),
\label{eq24}
\end{eqnarray}   
\noindent
where the func\-tion, $\Phi_0(z) = 3.5\times 10^{-3}\exp[-(z/1.7)^{1.47}]$ Mpc$^{-3}$ is given by Eq.~12 in Li et al~(2011) and $L_\star(z) = 1.4\times 10^{11} 10^{0.4[(z/1.78)^{0.47}]}$ L$_\odot$ is given by Eq. (A6) in Li et al~(2011). Note that the parameters describing the above relations have typical errors of the order of $\sim$10$\%$. We expect that their effect on the final spectra of neutrinos are much smaller than the effect of the unknown parameters describing the accretion and acceleration process of nuclei. Therefore, we do not consider such subtle effects during comparisons of the calculated neutrino spectra with the measured neutrino spectra since which have the uncertainties of the order of $\sim$2 (see spectral points in Fig.~3). In our calculations, we apply $\Omega_{\rm m} = 0.3$, 
$\Omega_\Lambda = 0.7$, and $H_0 = 70$ km s$^{-1}$ Mpc$^{-1}$. 
For the specific luminosity of spheroid galaxy, we derive the mass of the SMBH (from Eq.~23) and determine the accretion luminosity. 
The extragalactic neutrino background is calculated after integrating over the population of the 
SMBHs within the active galaxies and over different distances to the observer (up to $z_{\rm max} = 2$). The results are compared on Fig.~3, with the all-flavor ENB reported by the IceCube (Aartsen et al.~2015a), 
assuming different spectral indexes of the accelerated nuclei (on the left) and different values of the accretion efficiency onto the SMBHs (Fig.~3 on the right). The best description of the ENB is obtained for the spectral index equal to $\delta = 2.2$ and rather large accretion efficiency $\chi = 0.1$. The level of calculated ENB is consistent with the observations provided that the normalization factor of the spectrum of nuclei is equal to $A_{\rm N}\sim$2$\times 10^{-6}$. The factor $A_{\rm N}$ is the product of a few coefficients, i.e. the accretion power of the SMBH (in units of the Eddington luminosity) expressed by $\chi$, a factor describing the part of energy of nuclei taken by neutrons (in the range $\sim$8$^{-1}$ to $\sim$2$^{-1}$), the part of the accretion power transferred to the jet (assumed $2^{-1}$), the efficiency of acceleration of nuclei (usually assumed $10\%$), the parameter describing a part of all SMBHs which are active 
within galaxies, and the parameter describing a part of neutrons which reach the accretion disk.
The last two factors are the most uncertain. For the values of the other coefficients mentioned above the product of these two last parameters should be of the order of $(0.8-3.2)\times 10^{-3}$.    
Therefore, the model can explain the observations of the ENB provided that e.g. $(1.6-6.4)\%$ of the neutrons reach the accretion disk surface and $5\%$ of SMBHs are in the active phase.
In fact, the AGN fraction of field galaxies has been estimated on $5\%$ based on visual observations
(Dressler, Thompson \& Shectman~1985). Using the X-ray data for galaxy clusters with luminosity $\ge 10^{41}$ erg s$^{-1}$, Martini et al.~(2006) found the AGN fraction equal to $(5\pm 1.5)\%$, for galaxies magnitude brighter than $M_{\rm R} < -20$ (see also Arnold et al.~2009).
Moreover, the AGN fraction increases significantly for high-redshift clusters (e.g. Eastman et al.~2007).
We conclude that nuclei, accelerated and disintegrated in jets of SMBHs, can inject neutrons towards the accretion disks. They are able to produce very high energy neutrino background in the Universe recently observed by the IceCube.   

We have also investigated the effect of much lower ac\-cre\-tion ef\-fi\-ciency of mat\-ter onto the SMBHs by sho\-wing the neu\-trino spec\-tra for $\chi = 0.03$ and different spectral indexes of the neutrons
(Fig.~3 on the right). It is clear that the lower accretion efficiency might be also consistent with the observations provided that the spectral index of accelerated nuclei is clearly flatter than the previously considered value 2.2 (taking into account large error bars of the neutrino spectral points). We have also checked that the contribution to the ENB from SMBHs at larger red-shifts than 
$z = 2$ can be safely neglected.

\begin{figure*}[t]
\vskip 5.truecm
\includegraphics{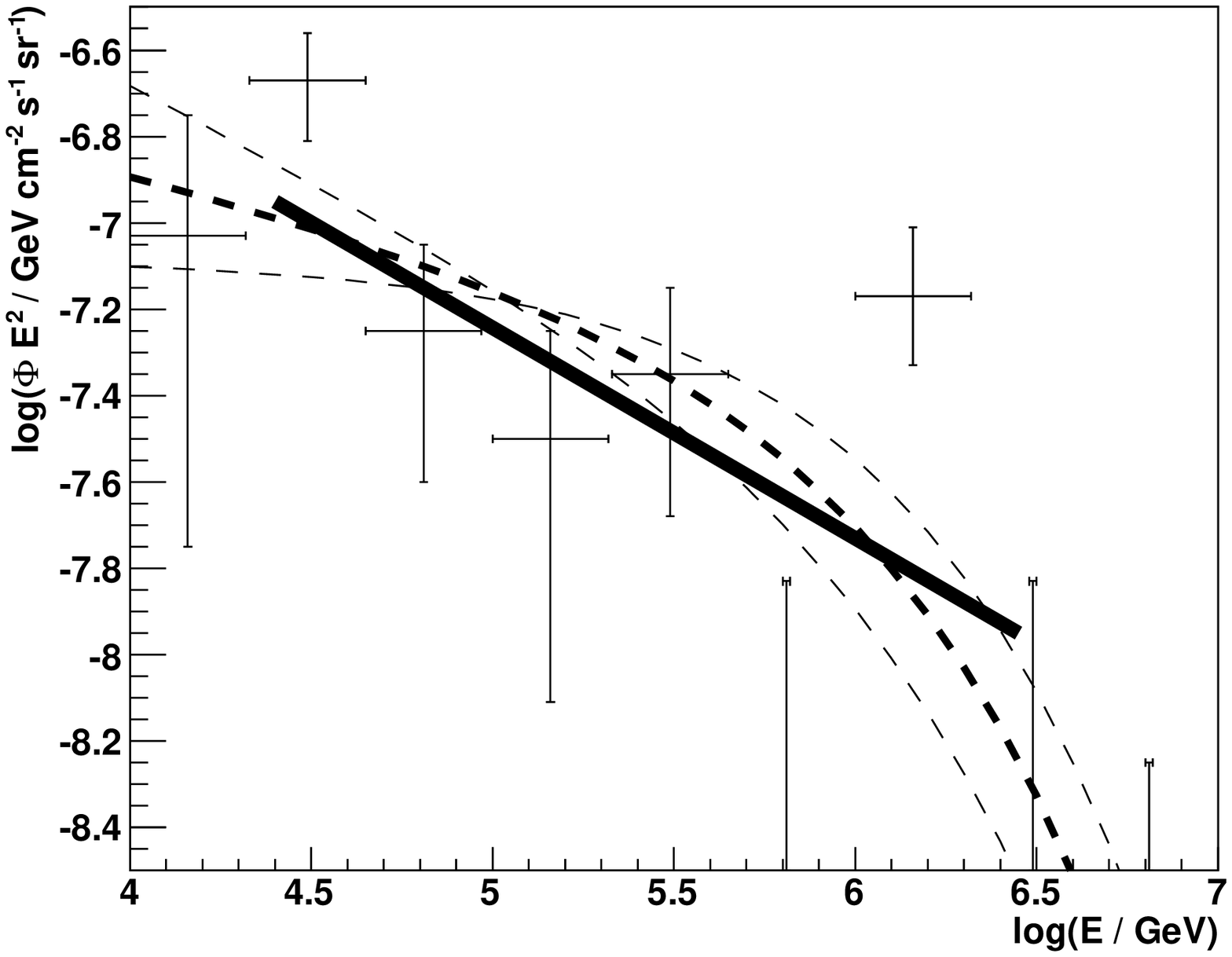}
\includegraphics{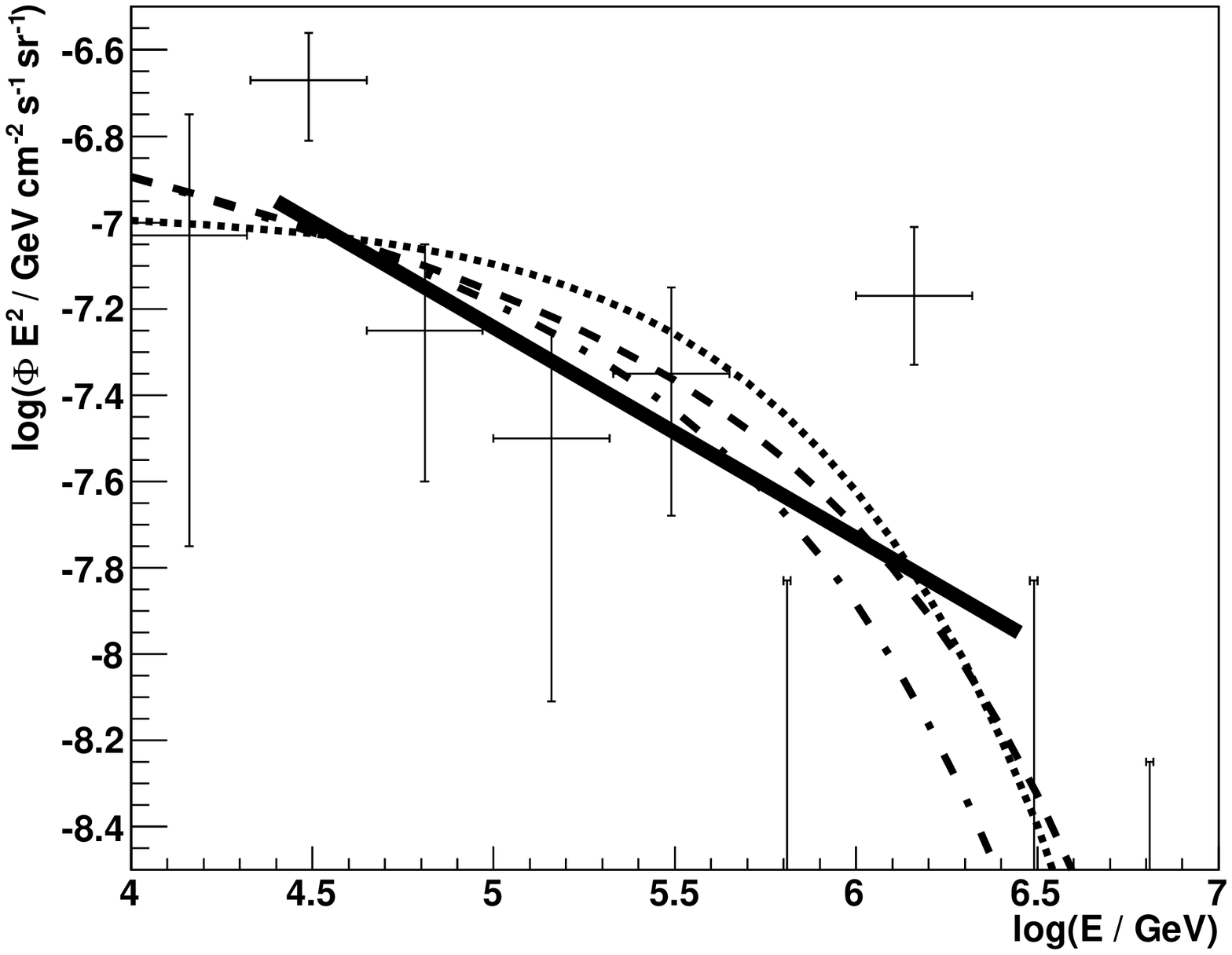}
\caption{Extragalactic (all flavor) Neutrino Background (ENB) produced by neutrons in collisions with the matter of the optically thick accretion disks around black holes in the Universe up to the redshift of $z_{\rm max} = 2$. The ENB, calculated for the power law spectrum of neutrons with the exponential cut-off at $E_{\rm n}^{\rm max}$ and spectral index equal to $\delta = 2.2$ (thick dashed curve), and 2. and 2.4 (thin dashed curves), is shown on the left figure. 
The normalization factor of the spectra (see the main text for precise definition) is equal to 
$A_{\rm N}\approx 2\times 10^{-6}$, $3\times 10^{-7}$, and $2\times 10^{-5}$, respectively for the spectral indexes mentioned above. The spectrum of neutrons extends above $\gamma_{\rm n}^{\rm min}$ (given by Eq.~21) and $\gamma_{\rm n}^{\rm max}$ is given by Eq.~20. The efficiency of the accretion process onto the SMBHs is fixed on $\chi = 0.1$. 
The population of the super-massive black holes in the Universe is modeled as described in Sect.~5. The dependence of the ENB on the efficiency of the accretion process, for $\chi = 0.1$ and $\delta = 2.2$ (dashed curve), $0.03$ and $\delta = 2$ (dotted, $A_{\rm N}\approx 10^{-6}$), $0.03$ and $\delta = 2.2$ (dot-dashed, $A_{\rm N}\approx 6\times 10^{-6}$), is shown on the right figure. The reported spectrum of (all flavor) neutrino background is shown by the solid error bars and its power law model by the thick solid line (Aartsen et al.~2015a).}
\label{fig3}
\end{figure*}

The neutrinos produced in terms of such a model are expected to be only mildly beamed in the direction
perpendicular to the accretion disk. Therefore, different types of observed active galaxies, i.e. blazars, radio galaxies, or even Seyfert galaxies with evidences of jets, could contribute to the ENB. Due to this feature of unbeamed neutrino emission, only the closest AGNs might be expected to produce observable neutrino event rates in the IceCube telescope.
In order to check this, we calculate expected neutrino event rates from the nearby active galaxy, 
M87, under the hypothesis that it is typical contributor to the ENB, i.e. assuming for it the average normalization factor $A_{\rm N} = 2\times 10^{-6}$ derived from the modeling of the ENB. For the active galaxy M87, which is at the distance of 16.4 Mpc (Bird et al. 2004) and harbors a SMBH with the mass $\sim$5$\times 10^9$ M$_\odot$ (Walsh et al.~2013), the expected neutrino event rate in the IceCube telescope is estimated on $\sim$0.7 $\nu$ yr$^{-1}$.
We used the effective area of the IceCube neutrino detector reported in Aartsen et al.~(2015b).
In these calculations we assumed the spectral index of nuclei equal to 2, i.e. marginally consistent with the modeling of the ENB (see Fig.~3). Note however, that M87 is under-luminous active galaxy 
with the estimated accretion rate $\dot{m}\approx 1.6\times 10^{-3}$ (di Matteo et al.~2003). The accretion process in M87 may not be correctly described by the considered in this paper Shakura-Sunyaev disk model. Therefore, this neutrino event rate should be considered as the upper limit.
We conclude that the perspectives for the identification of the neutrino events with the specific active galaxies are not very promising in terms of the considered model.

\section{Conclusion}

We propose that nuclei, accelerated in jets produced by the supermassive black holes in AGN, can be efficiently disintegrated in the interaction with the 
soft radiation from the inner part of the accretion disk. Neutrons, from their fragmentation, can reach dense regions of the accretion disk producing neutrinos in collisions with the disk matter.
For a typical temperature in the inner disk, the Lorentz factors of neutrons extracted from nuclei are clearly lower than those produced in pure proton-photon collisions (by one/two order of magnitudes). Therefore, the neutrino spectra extend down to the TeV energy range in contrast to the case of neutrons produced in the proton-photon collisions ($p-\gamma\rightarrow n + ...$), in which case the the neutrino spectrum is expected to flatten below PeV energies due to the higher threshold for the neutron production. Considered by us process of neutron production is also more efficient then proton-photon process due to the larger value of the cross section for the disintegration of nuclei.
Therefore, discussed here process of extraction of neutrons from nuclei seems to provide more suitable explanation of the observed cosmic neutrino events which show the power law spectrum clearly extending below a few PeVs down to $\sim$10 TeV.

We propose that nuclei can be accelerated in the magnetic field re-connection regions within the jet and/or at the boundary between fast jet and its slowly moving cocoon. Then, the basic parameters describing the model can be linked to the mass of the central super-massive black hole, the accretion rate onto this black hole and the magnetic field strength at the base of the jet.
Most of muons, produced in collisions of neutrons with the matter of the accretion disk, can decay before interacting since the density of matter in the disk is low. Therefore, the neutrino spectra can clearly extend from low energies up to the PeV energies.

We calculate the spectra expected from a specific SMBH within an active galaxy as a function of SMBH mass. SMBHs with larger masses are expected to produce neutrinos with larger fluxes extending to higher energies, provided that the accretion rate is constant (independent on the black hole mass) fraction of Eddington accretion rate. However, in order to produce significant fluxes of neutrinos at $\sim$PeV energies, the spectral index of primary nuclei should be close to 2. 

We aim to explain the recent measurements of the very high energy Extragalactic Neutrino Background
by the IceCube telescope (Aartsen et al.~2015a) in terms of our model. We determine the parameters of the population of the SMBHs in the Universe based on the observed link between masses of the SMBHs within the galaxies 
and their spheroid luminosity function. Good description of the ENB is obtained provided that the free parameter describing the neutrino flux produced in terms of our model is of the order of 2$\times 10^{-6}$. The parameter is the product of the fraction of active (accelerating nuclei) SMBHs, the accretion rate onto black hole, the acceleration efficiency of nuclei and the part of neutrons which reach the accretion disk. This factor seems to be reasonable. For example, it could be obtained provided that, AGN fraction of galaxies is $\sim 5\%$, the accretion rate is equal to $10\%$ of the Eddington rate, the nuclei are accelerated with $\sim 10\%$ efficiency and provided that $\sim$$(2-6)\%$ of neutrons reach the accretion disk. Moreover, the spectrum of accelerated nuclei should be of the power law type with the spectral index close to 2.2.

Since neutrinos, produced in such a model, are not strongly beamed along the jet axis, different types of active galaxies are expected to contribute to the observed ENB. In order to check whether such active galaxies might be directly observable by the IceCube neutrino telescope, we estimate the neutrino event rate expected from the nearby radio galaxy, M~87. It is assumed that M~87 is typical contributor to the ENB, i.e. the normalization of its emission is described by the factor obtained above from the fitting of the ENB. However, predicted neutrino event rate in the case of M87, $\sim 0.7$ neutrino events per year, should be considered as the upper limit since the accretion rate onto the SMBH in M87 is expected to be low.

I would like to thank the Referee for many useful comments and suggestions and J. Sitarek for reading the manuscript and comments. This work is supported by the grant through the Polish NCN No. 2014/15/B/ST9/04043.






\end{document}